\documentclass[twocolumn,superscriptaddress,floatfix,prl]{revtex4-1}
\usepackage{graphicx,amsfonts,amssymb,amsmath,hyperref,hypcap,enumerate}
\usepackage{color}
\usepackage{cleveref}

\newcommand{\qq}{\boldsymbol{q}}

\newcommand{\RR}{\boldsymbol{R}}
\newcommand{\rr}{\boldsymbol{r}}

\newcommand{\kk}{\boldsymbol{k}}

\newcommand{\bb}{\boldsymbol{b}}

\newcommand{\bv}{\boldsymbol{v}}

\begin{document}
\title{Theory of Phonon-Mediated Superconductivity in Twisted Bilayer Graphene}

\author{Fengcheng Wu}
\affiliation{Materials Science Division, Argonne National Laboratory, Argonne, Illinois 60439, USA}
\affiliation{Condensed Matter Theory Center and Joint Quantum Institute, Department of Physics, University of Maryland, College Park, Maryland 20742, USA}

\author{A. H. MacDonald}
\affiliation{Department of Physics, University of Texas at Austin, Austin, Texas 78712, USA}

\author{Ivar Martin}
\affiliation{Materials Science Division, Argonne National Laboratory, Argonne, Illinois 60439, USA}


\begin{abstract}
We present a theory of phonon-mediated superconductivity in near magic angle twisted bilayer graphene.  Using a microscopic model for phonon coupling to moir\'e band electrons, we find that phonons generate attractive interactions in both $s$ and $d$ wave pairing channels and that the attraction is strong enough to explain the experimental superconducting transition temperatures. Before including Coulomb repulsion, the $s$-wave channel is more favorable; however, on-site Coulomb repulsion can suppress $s$-wave pairing relative to $d$-wave. The pair amplitude varies spatially with the moir\'e period, and is identical in the two layers in the $s$-wave channel but phase shifted by $\pi$ in the $d$-wave channel.  We discuss experiments that can distinguish the two pairing states.
\end{abstract}

\maketitle

{\it Introduction.---}  Long-period moir\'e superlattices form whenever two-dimensional crystals are overlaid with a
small difference in lattice constant or orientation, and have recently been employed to
alter the electronic \cite{TBL2007,Hunt2013,Dean2013,Wang2015,Kim2017,Chen2018} and excitonic \cite{Wu2017, Wu2018, Yu2017} properties of two-dimensional materials.
One particularly exciting development is the
discovery
of interaction-induced insulating states accompanied by nearby lobes
of  superconductivity in twisted bilayer graphene \cite{Cao2018Magnetic,Cao2018Super}.
Both states occur only when the twist angle between layers is close to the largest magic angle \cite{Bistritzer2011} at which  the low-energy moir\'e bands \cite{Bistritzer2011,Morell2010,Mayou2012,Fang2016} are
nearly flat.  The superconducting pairing mechanism of
twisted bilayer graphene has not yet been established, and a variety of possibilities are
currently being explored \cite{Balents2018, Fu2018,Senthil2018, Baskaran2018, Padhi2018wigner,Dodaro12018, Liu2018chiral, Heikkila2018, Kang2018,Rademaker2018,Kennes2018strong, Isobe2018, Koshino2018, You2018},
including routes towards electron-electron interaction
mediated superconductivity, and also a phenomenological mean-field theory \cite{Heikkila2018} for $s$-wave pairing.  In this Letter we present a microscopic theory of magic angle superconductivity in twisted bilayer
graphene in which the attractive interaction is mediated by the phonon modes of the
individual two-dimensional graphene sheets. We find that phonons generate attraction in both $s$ and $d$ wave channels. In combination with the greatly enhanced density-of-states of the flat bands, the attraction in both channels is strong enough to account for the superconducting transition temperatures observed experimentally. The competition between $s$ and $d$ wave channels depends on Coulomb repulsive interaction. We discuss experimental signatures that could be used to distinguish these two pairing states.

\begin{figure*}[t]
    \includegraphics[width=1.8\columnwidth]{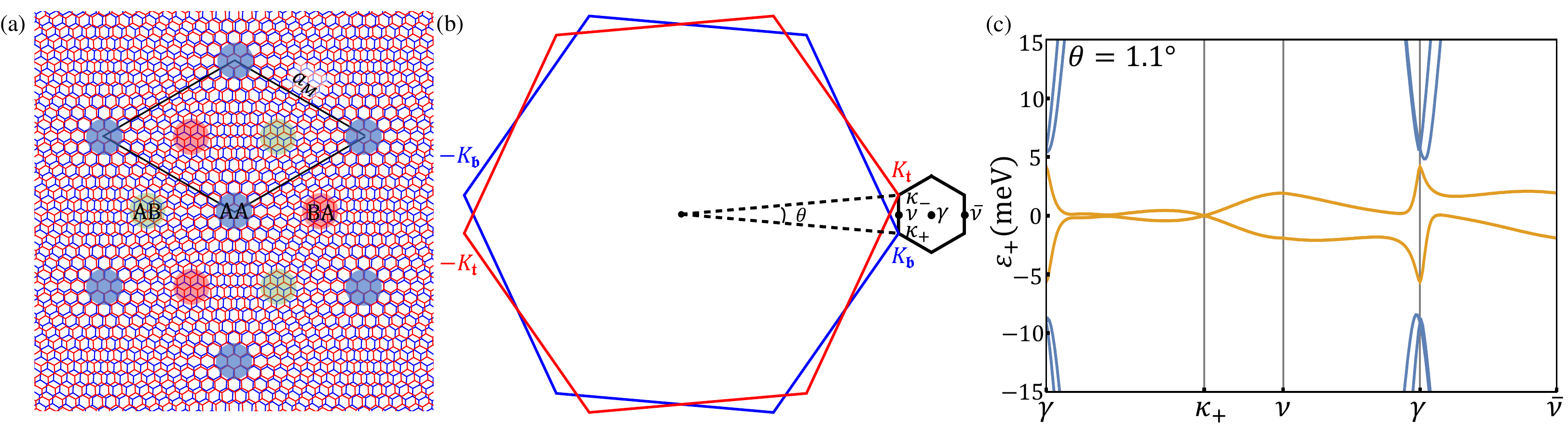}
	\caption{(a) Real  and (b) momentum space structure of twisted bilayer graphene. Low-energy electrons are located
	in $\pm K$ valleys.  (c) moir\'e bands in $+ K$ valley along high-symmetry lines at a twist close to the first magic angle. The spectra
    along $\nu \gamma$ and $\bar{\nu} \gamma$ lines are different,  reflecting the absence of time-reversal symmetry within one valley.}
	\label{Fig:band_structure}
\end{figure*}

{\it Moir\'e bands.---}
At small twist angles the single-particle physics of twisted bilayer graphene can be
described using a continuum moir\'e Hamiltonian, in which the atomic-scale commensurability plays no role.
To construct the twisted bilayer, we start from AA stacked bilayer graphene, and then
rotate the bottom and top layers by angles $-\theta/2$ and $+\theta/2$ around one of the hexagonal plaquette centers.  We choose the origin of coordinates to be on this
rotation axis and half-way between layers.
With respect to this origin, the point group symmetry is $D_6$, which is generated by a sixfold rotation $\hat{C}_6$
around the $\hat{z}$ axis, and twofold rotations  $\hat{\mathcal{M}}_x$ and
$\hat{\mathcal{M}}_y$ respectively around the $\hat{x}$ and $\hat{y}$ axes.
The operations  $\hat{\mathcal{M}}_{x, y}$ swap the two layers.
Because spin-orbit interactions
are negligible in graphene \cite{Brataas2006,Min2006},
electrons also have  accurate spin SU(2) symmetry
and spinless time-reversal symmetry $\hat{\mathcal{T}}$.

At small twist angles the bilayer moir\'e pattern, illustrated in Fig.~\ref{Fig:band_structure}(a),
is anchored by a triangular lattice of regions with local AA inter-layer coordination that has lattice constant
$a_M =  a_0/[2 \sin(\theta/2)]$, where $a_0$ is the lattice constant of monolayer graphene.
Local AB and BA coordination then occur at the corners of the moir\'e
Wigner-Seitz cell.  Our theory of phonon-mediated pairing
is based on the continuum moir\'e Hamiltonian
for low-energy electrons \cite{Bistritzer2011}, which is spin-independent and
is given in valley $+K$ by:
\begin{equation}
\mathcal{H}_{+}=\begin{pmatrix}
h_{\mathfrak{b}}(\kk) & T(\rr) \\
T^{\dagger}(\rr) & h_{\mathfrak{t}}(\kk)
\end{pmatrix}.
\label{Hmoire}
\end{equation}
Here $h_{\mathfrak{b, t}}$ are the isolated Dirac Hamiltonians of the
twisted bottom ($\mathfrak{b}$) and top ($\mathfrak{t}$) layers:
\begin{equation}
h_{\ell}(\kk) = e^{-i\ell \frac{\theta}{4} \sigma_z }[\hbar v_F (\kk-\boldsymbol{\kappa}_{\ell})\cdot \boldsymbol{\sigma}]e^{+i\ell \frac{\theta}{4} \sigma_z},
\end{equation}
where $\ell$ is $+1$ ($-1$) for the $\mathfrak{b}$ ($\mathfrak{t}$) layer,
$v_F$ is the bare Dirac velocity($\sim 10^6$ m/s), and
$\boldsymbol{\sigma}$ are Pauli matrices that act in the sublattice space.
Because of the rotation, the Dirac cone position in layer $\ell$ is shifted to
$\boldsymbol{\kappa}_{\ell}$.
We choose a moir\'e Brillouin zone (MBZ) in which the $\boldsymbol{\kappa}_{\ell}$ is
located at the  corners, and refer to the
MBZ center below as the $\gamma$ point.
The sublattice-dependent interlayer tunneling terms vary periodically with  the real space position $\rr$:
\begin{equation}
T(\rr)=w \Big[
T_0
+e^{-i \bb_+ \cdot \rr} T_{+1}
+e^{-i \bb_- \cdot \rr} T_{-1}
\Big]
\end{equation}
where $w \approx 118$ meV \cite{Jung2014} and
$T_j = \sigma_0 + \cos(2\pi j/3)\sigma_x+\sin(2\pi j/3)\sigma_y$.
We use $\bb$ to denote moir\'e reciprocal lattice vectors,
and $\bb_{\pm} = [4\pi/(\sqrt{3} a_M)](\pm1/2, \sqrt{3}/2)$.

Fig.~\ref{Fig:band_structure}(c) illustrates the $+K$-valley moir\'e band structure at a rotation angle that is close
to the largest magic angle.
The combined symmetry $\hat{C}_2 \hat{\mathcal{T}}$ implies that the Berry curvature of the moir\'e bands
is identically zero, and protects the Dirac cone band touching, and the $\hat{C}_3$ symmetry pins
the Dirac cones to the MBZ corners $\boldsymbol{\kappa}_{\ell}$.
Here $\hat{C}_2$ ($\hat{C}_3$) is a twofold  (threefold) rotation around $\hat{z}$ axis.
The $\hat{\mathcal{M}}_x$ operation maps $\boldsymbol{\kappa}_{+}$ to $\boldsymbol{\kappa}_{-}$, and therefore enforces 
 the energy spectra at $\boldsymbol{\kappa}_{\pm}$ to be identical.
The absence of
time-reversal symmetry in the single valley Hamiltonian implies that
$\varepsilon_{\tau}(\qq) \neq \varepsilon_{\tau}(-\qq)$, where $\tau=\pm$ labels valley
and $\qq$ is momentum relative to the $\gamma$ point.
Microscopic time-reversal invariance instead implies that $\varepsilon_{\tau}(\qq) = \varepsilon_{-\tau}(-\qq)$.
This feature in the single-particle band structure suggests that intra-valley electron pairing is not energetically favorable.
We therefore consider only inter-valley pairing in the following.

The velocity of the Dirac cones at $\boldsymbol{\kappa}_{\pm}$ varies systematically with twist
angle, decreasing from near $v_F$ at large twist angles and crossing through
zero at a series of magic twist angles. Near magic angle, the moir\'e band is nearly flat through
much of MBZ, leading to a greatly enhanced density-of-states and opening
the way to interaction driven phase transitions.
For our choice of parameters, the largest magic angle is $1^{\circ}$ if defined by vanishing Dirac velocity, while flat bands with the narrowest bandwidth ($\sim$3 meV) occur at about $1.05^{\circ}$.

\begin{figure}[b]
    \includegraphics[width=1\columnwidth]{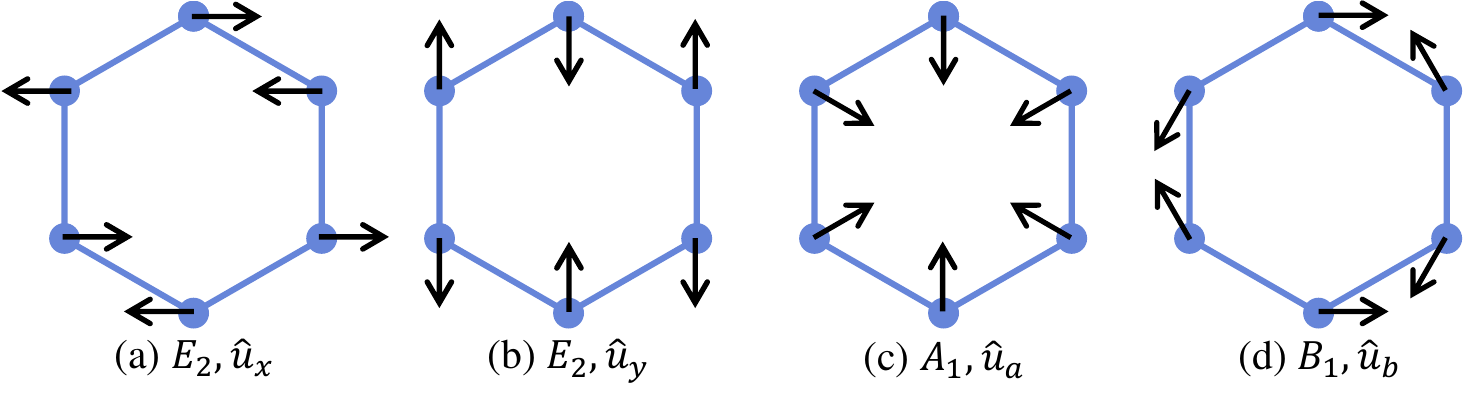}
	\caption{Illustration of the atomic displacements in the four phonon modes. Each blue dot represents a carbon atom.}
	\label{Fig:phonon}
\end{figure}

{\it Phonon mechanism.---}
We first study phonon-mediated pairing and then address effects of Coulomb repulsion in the discussion below.
We have considered a variety of phonon modes as discussed in the Supplemental Material \cite{SM}, and found that in-plane optical phonon modes associated with each graphene layer have a particularly strong effect on the pairing.
These modes yield weak phonon-mediated interlayer interactions \cite{Yan2008}, which we will neglect.
With this simplification we can follow the monolayer analysis in Ref.~\onlinecite{Aleiner2008}
which identifies four in-plane phonon modes that couple strongly to low-energy electrons:
(i) the doubly degenerate $E_2$ modes in the
vicinity of the $\Gamma$ point and (ii) the $A_1$ and $B_1$ modes,
which are combinations of phonon modes near $\pm K$.
The atomic displacements associated with the four modes are illustrated in Fig.~\ref{Fig:phonon}.
The $\Gamma$ and $\pm K$ point phonon modes lead respectively to intra and inter-valley scattering.
The isolated layer electron-phonon coupling Hamiltonian is:
\begin{equation}
\begin{aligned}
H_{\text{EPC}} = \int d^2 \rr \hat{\psi}^\dagger (\rr) \{
F_{E_2}[\hat{u}_x(\rr) \tau_z \sigma_y -\hat{u}_y (\rr) \sigma_x]\\
+F_{A_1} [\hat{u}_a(\rr) \tau_x \sigma_x + \hat{u}_b(\rr) \tau_y \sigma_x  ]
 \} \hat{\psi}(\rr),
\end{aligned}
\end{equation}
where $\hat{u}_{x,y}$ and $\hat{u}_{a, b}$ are the normal mode coordinates of the two $E_2$ modes, and
of the $A_1$ and $B_1$ modes respectively,  $\tau_{x, y, z}$ are Pauli matrices in valley space,
and the operator $\hat{\psi}$ is a spinor in sublattice-valley space:
$(\hat{\psi}_{+ A}, \hat{\psi}_{+ B}, \hat{\psi}_{- A}, \hat{\psi}_{- B})^{\text{T}}$.
(The subscripts $\pm$ refer to $\pm K$ valleys, $A$ and $B$ label sublattices, and the layer and spin indices are hidden.)
In a nearest-neighbor tight-binding model, the coupling constants
$F_{E_2}= F_{A_1} = (3/\sqrt{2})(\partial t_0 /\partial a_{CC}) $,
where $t_0$ and $a_{CC}$ are respectively the nearest-neighbor hopping parameter and distance.  When quantized the
$\hat{u}_\alpha$ can be expressed in terms of phonon creation and annihilation operators:
\begin{equation}
\begin{aligned}
&\hat{u}_\alpha (\rr) =\sqrt{\frac{\hbar}{2 N M \omega_\alpha}} \sum_{\qq} e^{i \qq \cdot \rr } [a_{\alpha} (\qq) + a_{\alpha}^{\dagger} (-\qq)],\\
\end{aligned}
\end{equation}
where $N$ is the number of $A$ sites in a monolayer, $M$ is the mass of a single carbon atom
and $\omega_\alpha$ is the frequency of phonon mode $\alpha$.
We neglect the momentum dependence of $\omega_\alpha$ below
because only phonons that are close to either $\Gamma$ or $\pm K$ points are important for low-energy electrons;
$\hbar \omega_{E_2}$ and $\hbar \omega_{A_1}$ are respectively
0.196 and 0.17 eV in monolayer graphene \cite{Aleiner2008}.
Integrating out the phonon modes and neglecting retardation effects because of the high phonon frequencies,
we obtain the following phonon-mediated interaction Hamiltonian:
\begin{equation}
\begin{aligned}
H_{\text{att}}=-\int d^2 \rr \{ g_{E_2}[(\hat{\psi}^\dagger \tau_z \sigma_y \hat{\psi})^2
+( \hat{\psi}^\dagger \sigma_x \hat{\psi})^2 ] \\
+g_{A_1} [ (\hat{\psi}^\dagger \tau_x \sigma_x \hat{\psi})^2
+ (\hat{\psi}^\dagger \tau_y \sigma_x \hat{\psi})^2 ] \}.
\end{aligned}
\label{Hatt}
\end{equation}
Here the operators $\hat{\psi}^\dagger$ and $\hat{\psi}$ are understood to be at the same coarse-grained position $\rr$,
and the attractive interaction strength $g_\alpha$ parameters are given by:
\begin{equation}
g_\alpha = \frac{\mathcal{A}}{N}\Big(\frac{F_\alpha}{\hbar \omega_\alpha}\Big)^2 \frac{\hbar^2}{2 M},
\end{equation}
where $\mathcal{A}$ is the sample area. We estimate $g_{E_2}$ and $g_{A_1}$ to be about  52 and 69 meV$\cdot$nm$^2$ respectively \cite{SM}.

To study the Cooper pairing instability, we restrict the interaction in (\ref{Hatt}) to the Bardeen-Cooper-Schrieffer
(BCS) channel that pairs electrons from opposite valleys:
\begin{equation}
\begin{aligned}
H_{\text{BCS}}=-4 \int d^2 \rr \{ g_{E_2}[\hat{\psi}^{\dagger}_{+ A s}\hat{\psi}^{\dagger}_{- A s'}\hat{\psi}_{- B s'}\hat{\psi}_{+ B s}+h.c.]\\+g_{A_1}[\hat{\psi}^{\dagger}_{+ A s}\hat{\psi}^{\dagger}_{- A s'}\hat{\psi}_{+ B s'}\hat{\psi}_{- B s}+ h.c. ]
\\+g_{A_1}[\hat{\psi}^{\dagger}_{+ A s}\hat{\psi}^{\dagger}_{- B s'}\hat{\psi}_{+ A s'}\hat{\psi}_{- B s}+ (A \leftrightarrow B) ]
\},
\end{aligned}
\label{HBCS}
\end{equation}
where $s$ and $s'$ are spin indices.
In $H_{\text{BCS}}$, there are two distinct spin-singlet pairing channels: (i) intra-sublattice pairing, e.g., $\epsilon_{s s' }\hat{\psi}^{\dagger}_{+ A s}\hat{\psi}^{\dagger}_{- A s'}$ and (ii) inter-sublattice pairing, e.g., $\epsilon_{s s' } \hat{\psi}^{\dagger}_{+ A s}\hat{\psi}^{\dagger}_{- B s'}$, where $\epsilon$ is the fully antisymmetric tensor with $\epsilon_{\uparrow \downarrow}=1$. 
Only the  $\pm K$ phonons contribute to inter-sublattice pairing, which has  $d$-wave symmetry because
electrons at different sublattices and opposite valleys share the {\it same} angular momentum under the threefold rotation
$\hat{C}_3 \hat{\psi}^{\dagger}(\rr) \hat{C}_3^{-1} = \exp[i 2\pi \sigma_z \tau_z /3] \hat{\psi}^{\dagger}(\mathcal{R}_3 \rr) $.  Inter-sublattice Cooper pairs therefore carry a finite angular momentum.  

\begin{figure}[t]
    \includegraphics[width=1\columnwidth]{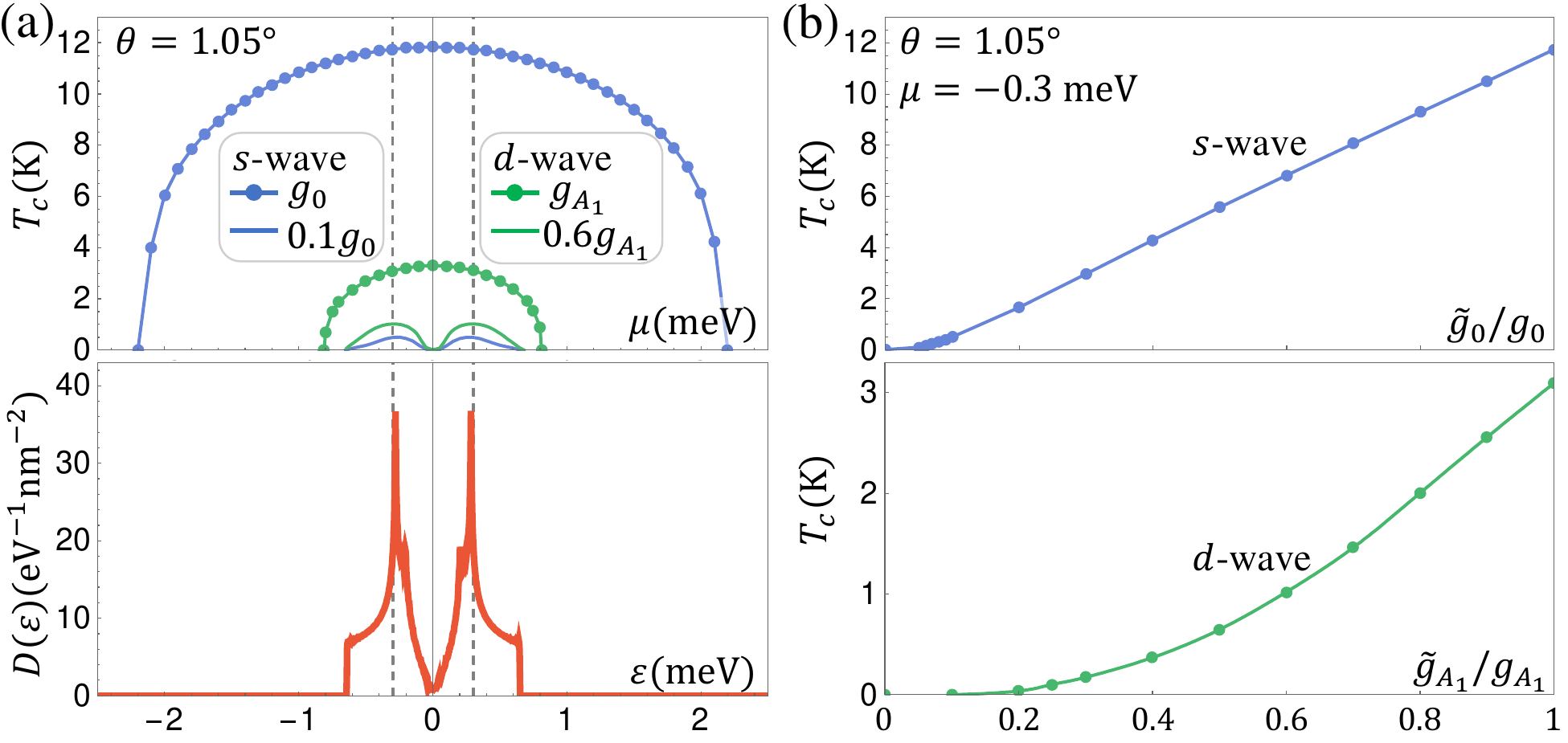}
	\caption{(a) Critical temperature $T_c$ in $s$-wave (blue lines) and $d$-wave channels (green lines)
	as a function of chemical potential (upper panel), and the DOS per spin and per valley as a function of
	energy (lower panel) for the twist angle $1.05^{\circ}$.
	The vertical dashed lines indicate the chemical potential at which the upper or lower flat band is half filled. (b) $T_c$ in $s$-wave (upper panel) and $d$-wave (lower panels) channels  as a function of reduced attractive interaction strength for $\theta=1.05^{\circ}$ and $\mu=-0.3 $ meV (half-filling of the lower flat band).}
	\label{Fig:Tc}
\end{figure}

{\it $s$-wave pairing.---}
In the $s$--wave intra-sublattice channel the local pairing amplitude,
\begin{equation}
\Delta_{\ell}^{(s)}(\rr) = \langle \hat{\psi}_{- \sigma \ell \downarrow}(\rr)  \hat{\psi}_{+ \sigma \ell \uparrow}(\rr) \rangle = - \langle \hat{\psi}_{- \sigma \ell \uparrow}(\rr)  \hat{\psi}_{+ \sigma \ell \downarrow}(\rr) \rangle,
\end{equation}
is sublattice ($\sigma$) independent by symmetry, but we allow a layer ($\ell$) dependence.
We solve the linearized gap equation by assuming that the pair amplitude has
the moir\'e periodicity and can therefore be expanded in the form
$\Delta_{\ell}^{(s)}(\rr) = \sum_{\bb} e^{i \bb \cdot \rr} \Delta_{\bb,\ell}^{(s)}$.  It follows that
\begin{equation}
\begin{aligned}
\Delta^{(s)}_{\bb,\ell} = & \sum_{\bb'\ell'} \chi_{\bb \bb'}^{\ell \ell'} \Delta^{(s)}_{\bb',\ell'}, \\
\chi_{\bb \bb'}^{\ell \ell'}= & \frac{ 2 g_0}{\mathcal{A}} \sum_{\qq,n_1,n_2}  \Big\{  \frac{1-n_F[\varepsilon_{n_1}(\qq)]-n_F[\varepsilon_{n_2}(\qq)]}{\varepsilon_{n_1}(\qq)+\varepsilon_{n_2}(\qq)-2\mu} \\
& \times [\langle u_{n_1}(\qq) | u_{n_2}(\qq) \rangle_{\bb, \ell}]^* \langle u_{n_1}(\qq) | u_{n_2}(\qq) \rangle_{\bb', \ell'} \Big\},
\end{aligned}
\label{chi}
\end{equation}
where $\chi$ is the pair susceptibility, $g_0 = g_{E_2}+g_{A_1}$,
$\qq$ is a momentum within MBZ,
$n_{1, 2}$ are moir\'e band labels in $+K$ valley,
$\varepsilon_n$ and $|u_n\rangle$ are the corresponding energies and wave functions,
$n_F(\varepsilon)$ is the Fermi-Dirac occupation function, and $\mu$ is the chemical potential.
The overlap function $\langle ... \rangle_{\bb, \ell}$ is the layer-resolved matrix element of the plane-wave operator $\exp(i \bb \cdot \rr )$. Note that $\hat{\mathcal{T}}$ symmetry has been employed to simplify (\ref{chi}).

\begin{figure}[t]
    \includegraphics[width=1\columnwidth]{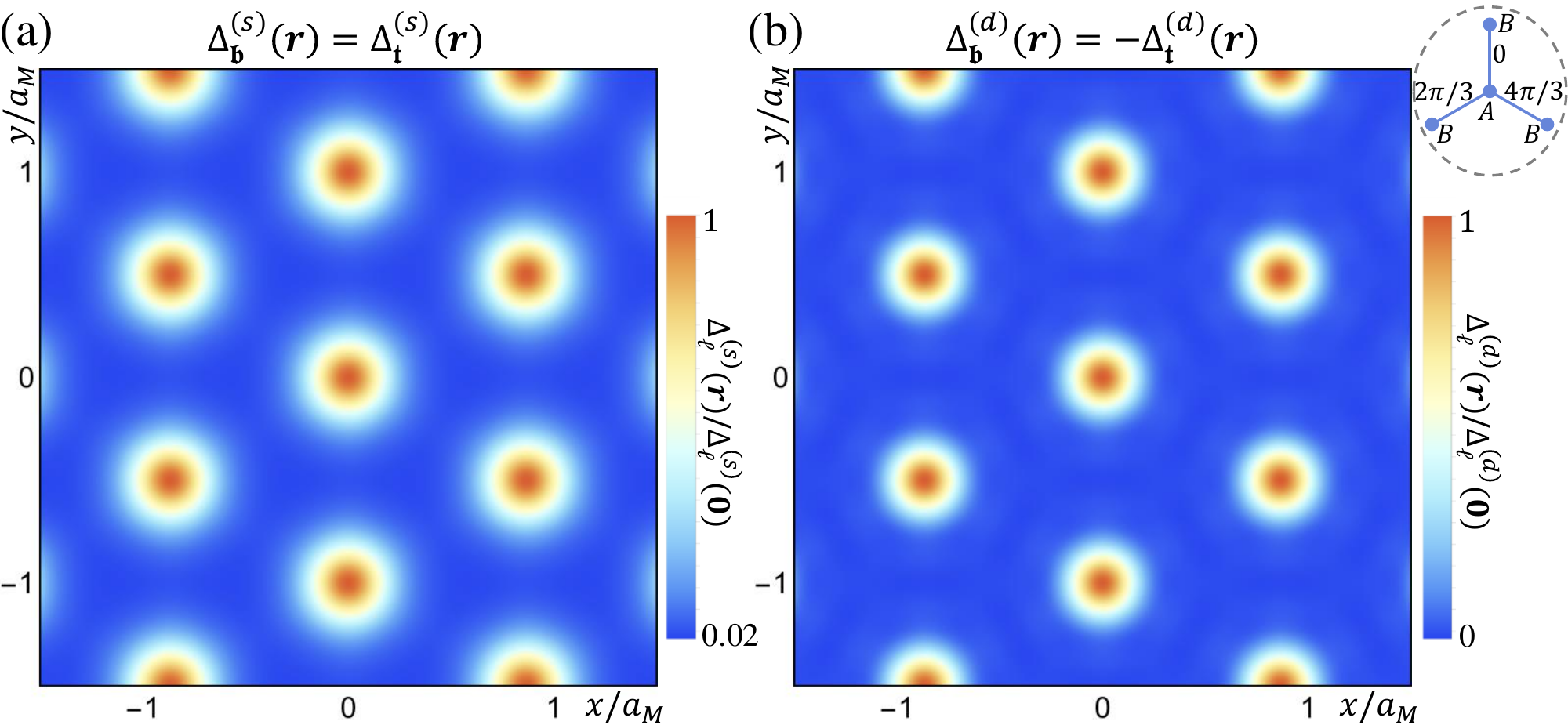}
	\caption{Real-space map of pair amplitudes $\Delta(\rr)$ that satisfy
	 (a) $s$-wave and (b) $d$-wave linearized gap equations for $\theta=1.05^{\circ}$ and $\mu=-0.3$ meV.
	 $\Delta(\rr)$ is peaked in the AA regions in both cases.
	 The inset in (b) illustrates $d_+$  pairing at the atomic scale, where electrons in the same layer but
	 on different sublattices are paired with the indicated bond-dependent phase factors.}
	\label{Fig:sc_potential}
\end{figure}

The critical temperature $T_c$ is reached when the largest eigenvalue of $\chi$ is equal to 1.
In  Fig.~\ref{Fig:Tc}(a) we illustrate $T_c$ calculated in this way for $\theta=1.05^{\circ}$,
including momenta $\bb$ up to the third moir\'e reciprocal lattice vector shell.
The relatively large $T_c$  values, which exceed $10$ K near the magic angle,
can be understood by examining the uniform susceptibility,
which has the standard $g_0 \int d \varepsilon D(\varepsilon) [1-2 n_F(\varepsilon)]/[2(\varepsilon-\mu)]$ form,
where $D(\varepsilon)$ is the density of states (DOS) per spin and per valley.
Because the DOS can reach values of order $10$ eV$^{-1}$ nm$^{-2}$,
as illustrated in Fig.~\ref{Fig:Tc},
the coupling constant $g_0 D(\mu)$ can be of order one, corresponding to
strong coupling superconductivity.
Since the higher-energy bands have much smaller DOS, we have retained only the two
flat bands in evaluating (\ref{chi}).
The eigenvector of $\chi$ with the largest eigenvalue specifies the spatial and
layer dependence of the pair amplitude.  For $s$-wave pairing, the pair amplitude is layer independent
and concentrated near AA regions in the moir\'e pattern, as illustrated Fig.~\ref{Fig:sc_potential}(a).
The spatial variation of $\Delta^{(s)}$ follows the electron density distribution in the normal state of the flat bands.
$\Delta^{(s)}$ transforms trivially under all the point-group
symmetries, confirming that intra-sublattice pairing is $s$-wave.

{\it $d$-wave pairing.---}
In the inter-sublattice channel, the pair amplitudes
$\epsilon_{s s' } \hat{\psi}^{\dagger}_{+ A s}\hat{\psi}^{\dagger}_{- B s'}$
and  $\epsilon_{s s' } \hat{\psi}^{\dagger}_{+ B s}\hat{\psi}^{\dagger}_{- A s'}$
carry opposite angular momenta corresponding to {\it chiral} $d$-wave pairing channels
which we refer to as
$d_+$ and $d_-$ respectively.  At the atomic scale, chiral $d$-wave pairing is realized by forming
nearest-neighbor spin-singlet Cooper pairs with bond-dependent phase factors, as illustrated in
Fig.~\ref{Fig:sc_potential}(b). 
Because of $\hat{\mathcal{T}}$ symmetry, both channels
have the same $T_c$ and we therefore focus on $d_+$ pairing,
which has the same susceptibility as in Eq.~(\ref{chi})
except that (i) $g_0$ is replaced by $2g_{A_1}$ and (ii) the overlap functions are
replaced by $\langle u_{n_1}(\qq) |\sigma_+| u_{n_2}(\qq) \rangle_{\bb, \ell}$, where
$\sigma_+ = (\sigma_x+ i \sigma_y)/2$.
The operator $\sigma_+$ is closely related to the velocity operator
$\hat{v}_x+i \hat{v}_y$, where $\hbar  \hat{\bv}=\partial \mathcal{H}_+/\partial \kk $.
Near the magic angle, the velocity of the flat bands is strongly suppressed, but
the layer counter-flow velocity remains large \cite{Bistritzer2011}.
As a result, we find that the leading $d$-wave instability has pair amplitudes of opposite signs in the two layers:
$\Delta_{\mathfrak{b}}^{(d)}(\rr) = - \Delta_{\mathfrak{t}}^{(d)}(\rr)$,
where $\Delta_{\ell}^{(d)}=\langle \hat{\psi}_{- B \ell \downarrow} \hat{\psi}_{+ A \ell \uparrow} \rangle$.
The highest $T_c$ in $d$-wave channel at $\theta = 1.05^{\circ}$ is about 3 K, as shown in Fig.~\ref{Fig:Tc}(a).
The spatial variation of the pair amplitude, illustrated in Fig.~\ref{Fig:sc_potential}(b), is similar to the $s$-wave case.
We note that $\Delta_{\ell}^{(d)}(\rr)$ describes the
center-of-mass motion of the Cooper pairs, while the relative motion of the two paired electrons
has $d$-wave symmetry.

{\it Discussion.---}
Experimentally, magic angle twisted bilayer graphene
exhibits superconducting states when the lower energy flat band is near half filled, but
not at neutrality ($\mu=0$) or when the upper  flat band is partially occupied \cite{Cao2018Super}.  
The highest experimental $T_c$ is $\sim 1.7$ K.
It is not yet clear whether or not this stark
particle-hole asymmetry, which is absent in our results, is intrinsic or due
to an extrinsic disorder effect.
We note that the moir\'e band structure does have intrinsic particle-hole asymmetry as evident in Fig.~\ref{Fig:band_structure}(c), and the asymmetry can be sensitive to the exact model parameters. 
Even discounting the particle-hole asymmetry, differences remain between experimental and theoretical
$T_c(\mu)$ trends in Fig.~\ref{Fig:Tc}, particularly in connection with the
experimental absence of superconductivity at neutrality.
Adding a Coulomb contribution to
electron-electron interactions weakens
the attractive interaction and makes the $T_c$
calculation more like the standard weak-coupling case
in which only $g_0 D(\mu)$ is relevant, and could explain the absence
of superconductivity at neutrality where $D(\mu)$ has a minimum and often vanishes, as shown in Fig.~\ref{Fig:Tc}(a).

We account for the Coulomb repulsion using a phenomenological model with only the atomic-scale on-site ($U_0$) and nearest-neighbor ($U_1$) repulsion on the honeycomb lattice of each graphene layer. $U_0$ and $U_1$ respectively suppress $s$ and $d$ wave pairings \cite{SM}, and the corresponding attractive interaction strengths in the gap equation are reduced to $\tilde{g}_0 = g_0 - U_0 \mathcal{A}_0 /2$ and $\tilde{g}_{A_1} = g_{A_1} - 3 U_1 \mathcal{A}_0 /4$, where $\mathcal{A}_0 = \sqrt{3} a_0^2/2$. We have calculated $T_c$ as a function of $\tilde{g}_0$ and $\tilde{g}_{A_1}$ respectively for the two channels [Fig.~\ref{Fig:Tc}(b)]. If $T_c$ in $s$-wave channel is fit to the experimental value 1.7 K for a half-filled lower flat band at $\theta=1.05^{\circ}$, then $U_0$ is about 3.7 eV. Similarly, $U_1$ is about 0.5 eV if $d$-wave has $T_c\sim $ 1.7 K. Depending on the exact values of $U_{0, 1}$, either channel can be the leading superconductivity instability. We note that the on-site repulsion $U_0$ can drive correlated insulating states for integer number of electrons or holes per moir\'e cell, which has been studied for the charge neutral case \cite{AFMTBLG}.
Experimentally, the interaction-induced insulating states at half filling of the lower or upper flat band have a tiny gap of about 0.3 meV \cite{Cao2018Magnetic,Cao2018Super}. This tiny gap is a possible indication that the Coulomb repulsion is strongly screened, which can be due to the enhanced DOS in the moir\'e bands, the dielectric encapsulation (hexagonal boron nitride) and the nearby metallic gate that is only 10$\sim$30 nm away from the twisted bilayer \cite{Cao2018Magnetic,Cao2018Super}. The screening effects require further study, while the phonon-mediated attraction combined with the local repulsion can provide a minimal model to study the competition between the superconducting and the correlated insulating states. Our theory should be taken as a step toward a full quantitative theory of twisted bilayer graphene.

If $d$-wave is the leading instability, the $d_{\pm}$ pair amplitudes have identical $T_c$, and the
corresponding superconducting state then has a two-component order parameter that can lead to
either chiral or nematic superconductivity.
The chiral state is more favored in mean-field theory because it is fully gapped, whereas
nematic pairing with spontaneous rotational symmetry breaking results in point nodes. In a follow-up work, we will report that the chiral $d$-wave state is topological and carries spontaneous bulk supercurrent. 

The $s$ and $d$ wave pairings with distinct in-plane phase structures can be distinguished by phase-sensitive experiments \cite{vanHarlingen1995}. In addition, a peculiar feature of the proposed $d$-wave superconductivity is that the order parameter has opposite signs on the top and bottom graphene layers, which could also be tested by phase-sensitive measurement. The chiral $d$-wave state spontaneously breaks time reversal symmetry, which can be examined by Kerr effect and magnetization measurement. Moreover, application of uniaxial strain, which tunes the competition between nematic and chiral $d$-wave states, in conjunction with upper critical field or critical current measurements could be used to differentiate $s$ and $d$ wave pairings.

Our theory sets up a general framework to study superconductivity in twisted bilayer graphene, and can be used to predict the response of superconductivity to various perturbations, such as  electric displacement field, magnetic field, pressure and disorder, which can be compared with coming experimental results. 

{\it Acknowledgment.---}
We thank M. Norman for valuable discussions. Work at Argonne was supported by the Department of Energy, Office of Science,
Materials Science and Engineering Division. F. W. was also supported by Laboratory for Physical Sciences.
A. H. M. was supported by the Department of Energy, Office of Basic Energy Sciences
under  grant DE-FG02-02ER45958 and by the Welch foundation under grant TBF1473.

{\it Note added.} A new experimental work has recently been posted \cite{Dean2018Tuning}, which demonstrates the tunability of the superconductivity by pressure and also reports the appearance of superconductivity in partially filled upper flat bands.

\bibliographystyle{apsrev4-1}
\bibliography{refs}

\clearpage
\begin{center}
	\textbf{Supplemental Material}
\end{center}

This Supplemental Material includes the following four sections: (1) the electron-phonon coupling generated by different phonon modes; (2) Coulomb repulsion effects on the pairing; (3) twist-angle dependence of the superconductivity critical temperature; (4) discussion on the coherence length and the spatial variation of the pair amplitude.

\section{Electron-phonon coupling}
We provide additional discussion on electron-phonon coupling. To estimate the coupling constants $F_{E_2}$ and $F_{A_1}$, we used the following hopping function:
\begin{equation}
t_0 = \tilde{t}_0 \exp[-\beta (\frac{a_{CC}}{\tilde{a}_{CC}}-1)],
\end{equation}
where $\tilde{t}_0 = -3$ eV is the static nearest-neighbor hopping parameter in monolayer graphene, and $\tilde{a}_{CC}$ is the equilibrium nearest-neighbor carbon distance. We take the exponential decay factor $\beta$ to be 3.3 \cite{Ribeiro2009}. The coupling constants are given by:
\begin{equation}
\begin{aligned}
F_{E_2}=& F_{A_1} = \frac{3}{\sqrt{2}}\frac{\partial t_0} {\partial a_{CC}}\\
=&-\frac{3\beta}{\sqrt{2}} \frac{\tilde{t}_0}{\tilde{a}_{CC}}=-3\beta \sqrt{\frac{3}{2}} \frac{\tilde{t}_0}{a_0},
\end{aligned}
\end{equation}
where $a_0$ is the lattice constant of monolayer graphene. The attractive interaction strength is estimated as follows:
\begin{equation}
\begin{aligned}
g_\alpha & = \frac{\mathcal{A}}{N}\Big(\frac{F_\alpha}{\hbar \omega_\alpha}\Big)^2 \frac{\hbar^2}{2 M}\\
&= \frac{\sqrt{3}}{2} a_0^2 \Big(\frac{F_\alpha}{\hbar \omega_\alpha}\Big)^2 \frac{\hbar^2}{2 M} \\
&= \frac{27 \sqrt{3}}{4} \beta^2 \Big(\frac{t_0}{\hbar \omega_\alpha} \Big)^2 \frac{\hbar^2}{2 M},
\end{aligned}
\label{galpha}
\end{equation}
and the numerical values of $g_{E_2}$ and $g_{A_1}$ are respectively about 52 and 69 meV$\cdot$nm$^2$.

The $E_2$, $A_1$ and $B_1$ modes have in-plane atomic displacements, and their coupling to the interlayer tunneling should be weak because (1) the interlayer coupling strength $w$ is an order of magnitude smaller compared to the in-plane hopping parameter $t_0$ and (2) the in-plane displacements are ineffective in changing the out-of-plane bond length. Therefore, we neglect the coupling between in-plane phonon modes and interlayer tunneling.

The layer breathing mode, in which the two layer move relative to each other in $\hat{z}$ direction, can couple to the interlayer tunneling. The attractive interaction constant generated by such phonon mode will be similar to (\ref{galpha}) with the major modification that $(t_0/\hbar \omega_\alpha)^2$ is replaced by $(w/\hbar \omega_z)^2$, where $\hbar \omega_z \approx 11 $ meV is the frequency of the layer breathing mode \cite{Yan2008}. Although $(w/\hbar \omega_z)^2 \approx 0.5 (t_0/\hbar \omega_{E_2})^2$, we find that the coupling to the layer breathing mode is strongly suppressed near the magic angle for the following reason. The coupling is generated by the tunneling matrix $T(\rr)$ as defined in the main text. When the twist angle approaches the magic angle, the average value of the tunneling terms with respect to the flat bands becomes very small in magnitude. This is consistent with the fact that flat band energy becomes nearly zero (Dirac point energy is set as the energy zero), and the tunneling terms are just part of the Hamiltonian that contributes to the total energy. 

The layer shear mode should also have weak coupling to electrons. In the shear mode, the two layer move relative to each other in $\hat{x}$ and $\hat{y}$ directions but there is no relative motion between two sublattices within the same layer. The rigid shear mode with zero momentum does not couple to the intralayer hopping. Its coupling to interlayer tunneling also effectively vanishes when there is a twist between layers, because a relative displacement between layers will just lead to a global shift of the moir\'e pattern and have no influence on the electronic energy spectrum \cite{Bistritzer2011}.

The coupling between acoustic phonon modes and electrons is proportional to Fermi wave vector $k_F$, which is small because the moir\'e period $a_M$ is much greater than the monolayer lattice constant $a_0$. 

Thus, we do not take into account the layer breathing modes, shear modes and acoustic modes for phonon mediated superconductivity in this work, and leave a detailed investigation of these phonon modes to future work.

In theories of phonon-mediated superconductivity, vertex corrections to the electron-phonon interaction can usually be neglected because of the
large mismatch between Fermi and sound velocities, as stated in Migdal's theorem.  This conclusion  applies equally well to two-dimensional Dirac systems \cite{Roy2014}.
In an isolated graphene sheet the ratio of the Dirac velocity to the sound velocity is greater than 10$^2$.  Since the Dirac velocity in a twisted graphene bilayer is suppressed by factors this large only in extremely narrow ranges of
twist angle, we expect that although vertex corrections may affect
near-magic-angle superconductivity quantitatively, our results, which neglect their effects, should remain qualitatively valid.

\section{Coulomb Repulsion}
In the main text, we have considered a phenomenological model for the Coulomb interaction that only retains the on-site ($U_0$) and nearest-neighbor ($U_1$) repulsion on the honeycomb lattice of each graphene layer. Here we show how to transform the interaction strength of a lattice model to that of a continuum model. 

The on-site repulsion on a monolayer graphene honeycomb lattice is described by
\begin{equation}
\begin{aligned}
H_0 &= U_0 \sum_{\RR, \sigma} c^{\dagger}_{\RR\sigma \uparrow}c^{\dagger}_{\RR\sigma \downarrow}c_{\RR\sigma\downarrow}c_{\RR\sigma\uparrow}\\
&=\frac{U_0}{N}\sum_{\kk_i}^{'} \sum_{\sigma}  c^{\dagger}_{\kk_1\sigma \uparrow}c^{\dagger}_{\kk_2\sigma \downarrow}c_{\kk_3\sigma\downarrow}c_{\kk_4\sigma\uparrow},
\end{aligned}
\label{H0}
\end{equation}
where $\sigma$ labels the $A$ and $B$ sublattices, $N$ is the number of $A$ sites in the monolayer, and the prime on the summation of the second line enforces the momentum conservation $\kk_1+\kk_2=\kk_3+\kk_4$. To obtain a continuum model, we retain low-energy states in $\pm K$ valleys:
\begin{equation}
H_0\approx \frac{U_0}{N}
\sum_{\kk_i}^{'} \sum_{\tau_i}^{'}\sum_{\sigma}
c^{\dagger}_{\kk_1 \tau_1 \sigma \uparrow}c^{\dagger}_{\kk_2 \tau_2 \sigma \downarrow}c_{\kk_3 \tau_3 \sigma\downarrow}c_{\kk_4 \tau_4 \sigma\uparrow},
\end{equation}
where $\tau = \pm 1$ is the valley index, and the prime on the summation of $\tau_i$ implies the valley conservation $\tau_1+\tau_2=\tau_3+\tau_4$. In the operator $c^{\dagger}_{\kk \tau \sigma s}$, the momentum $\kk$ is measured relative to $\tau K$. We make a Fourier transformation to introduce the coarse-grained real-space position $\rr$:
\begin{equation}
c^{\dagger}_{\kk \tau \sigma s} = \frac{1}{ \sqrt{\mathcal{A}} }\int d\rr e^{i \kk\cdot\rr} \hat{\psi}^{\dagger}_{\tau \sigma s}(\rr),
\end{equation}
where $\mathcal{A}$ is the system area.
The on-site repulsion can then be transformed to a continuum model with a delta-function interaction:
\begin{equation}
H_0 \approx U_0 \mathcal{A}_0 \sum_{\tau_i}^{'} \sum_{\sigma} \int d\rr 
\hat{\psi}^{\dagger}_{\tau_1 \sigma \uparrow}
\hat{\psi}^{\dagger}_{\tau_2 \sigma \downarrow}
\hat{\psi}_{\tau_3 \sigma \downarrow}
\hat{\psi}_{\tau_4 \sigma \uparrow},
\label{H_delta}
\end{equation}
where $\mathcal{A}_0 = \sqrt{3} a_0^2 /2$ is the area per $A$ site, and operators $\hat{\psi}^\dagger$ and $\hat{\psi}$ are at the same position $\rr$. By comparing $H_0$ with Eq.~(8) of the main text, we find that the on-site repulsion only suppresses $s$-wave pairing, and the corresponding attractive interaction strength in the gap equation is reduced to $\tilde{g}_0=g_0-U_0 \mathcal{A}_0/2$.

We carry out similar analysis for nearest-neighbor repulsion:
\begin{equation}
\begin{aligned}
H_1 &= U_1 \sum_{\RR, \boldsymbol{\delta}, s,s'} 
c^{\dagger}_{\RR A s}
c^{\dagger}_{(\RR+\boldsymbol{\delta}) B s'}
c_{(\RR+\boldsymbol{\delta}) B s'}
c_{\RR A s}\\
&=\frac{U_1}{N} \sum_{\kk_i}^{'}\sum_{s,s'} 
\mathcal{F}(\kk_3-\kk_2)  c^{\dagger}_{\kk_1 A s}c^{\dagger}_{\kk_2 B s'}c_{\kk_3 B s'}c_{\kk_4 A s}
\end{aligned}
\end{equation}
where $\boldsymbol{\delta}$ denotes the three bond vectors that connect nearest neighbors on the honeycomb lattice, and the form factor is given by $\mathcal{F}(\kk)=\sum_{\boldsymbol{\delta}} e^{i\kk\cdot \boldsymbol{\delta}}$. When states are restricted to $\pm K$ valleys, the form factor $\mathcal{F}(\kk_3-\kk_2)$ suppresses inter-valley scattering, which we neglect. This leads to the continuum Hamiltonian:
\begin{equation}
\begin{aligned}
H_1 & \approx  \frac{3U_1}{N}
\sum_{\kk_i}^{'} \sum_{\tau, \tau'}\sum_{s, s'}   
c^{\dagger}_{\kk_1 \tau A s}c^{\dagger}_{\kk_2 \tau' B s'}c_{\kk_3 \tau' B s'}c_{\kk_4 \tau A s}\\
& = 3U_1 \mathcal{A}_0 \sum_{\tau, \tau'}\sum_{s, s'} 
\int d\rr
\hat{\psi}^{\dagger}_{\tau A s}
\hat{\psi}^{\dagger}_{\tau' B s'}
\hat{\psi}_{\tau' B s'}
\hat{\psi}_{\tau A s},
\end{aligned}
\end{equation}
which suppresses $d$-wave pairing, and its attractive interaction strength in the gap equation is reduced to $\tilde{g}_{A_1}=g_{A_1}-3U_1 \mathcal{A}_0/4$.

In the main text, we made an estimation of $U_0$ and $U_1$ in order to fit with the experimental $T_c$ value, which leads to $U_0 \approx 3.7$ eV for $s$-wave channel and $U_1 \approx 0.5$ eV for $d$-wave channel. In free-standing monolayer graphene, $U_0$ and $U_1$ are estimated to be 9.3 eV and 5.5 eV based on the constrained random phase approximation \cite{GrapheneU0}. Screening effects in doped twisted bilayer graphene should be stronger compared to pristine monolayer graphene due to the greatly enhanced DOS. Furthermore, the dielectric encapsulation (hexagonal boron nitride with a dielectric constant of about 5) and the nearby metallic gates used in the experiments should also contribute to screening, although the on-site Coulomb interaction might be less affected by the dielectric environment. For $U_1$, an effective dielectric constant of 11 is required to reduce the value 5.5 eV in pristine graphene to 0.5 eV, which is possible given the above intrinsic high DOS and extrinsic (dielectric environment) screening effects.

Because of the high DOS when the flat bands are partially filled, the Coulomb interaction in real space is strongly suppressed for distance greater than $a_M$. The screening effects of Coulomb repulsion over distance shorter than $a_M$ requires a calculation of the momentum dependent dielectric function, which we leave for future work. On the other hand, the phonon-mediated attraction combined with the local repulsion can provide a minimal model to study the competition between the superconducting and the correlated insulating states. In the experiment \cite{Cao2018Super}, the transition temperature for the correlated insulating states ($T_c \sim$ 4 K) and superconductors ($T_c \sim$ 2 K) are comparable. Our interpretation is that the repulsion responsible for the correlated insulating states and the attraction responsible for the superconductors have comparable strength, and our estimation is consistent with this scenario.

\section{Twist angle dependence of $T_c$}
\begin{figure}[t]
	\includegraphics[width=1\columnwidth]{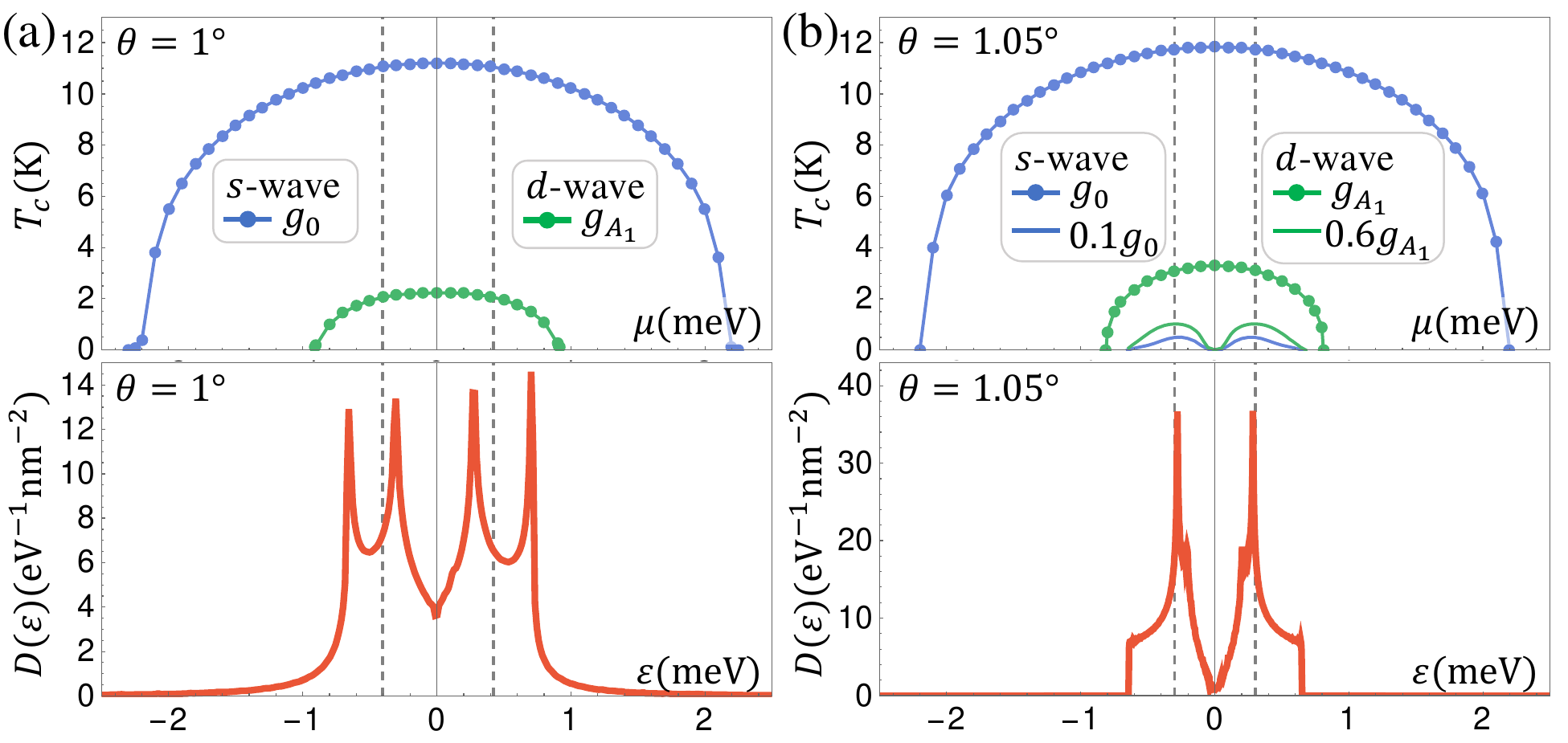}
	\caption{Critical temperature $T_c$ in $s$-wave (blue lines) and $d$-wave channels (green lines)
		as a function of chemical potential (upper panel), and the DOS per spin and per valley as a function of
		energy (lower panel) for the twist angle (a) $1^{\circ}$ and (b) $1.05^{\circ}$.
		The vertical dashed lines indicate the chemical potential at which the upper or lower flat band is half filled.}
	\label{Fig:Tc_1}
\end{figure}

We find that the strongly enhanced DOS near magic angle is crucial for the superconductivity. This is consistent with the experimental fact that superconductivity only occurs within a narrow range of the twist angle. To obtain critical temperature of few Kelvins, the DOS per spin and per valley should be on the order of 5 eV$^{-1}$ nm$^{-2}$ given the attractive interaction strength generated by phonons. In Fig.~\ref{Fig:Tc_1}, we show the critical temperature for a slightly different angle $1^{\circ}$, which has a smaller DOS compared to $1.05^{\circ}$ and the $T_c$ is correspondingly lower.

In main text, we have used the extremely narrow bands (bandwidth about 3 meV at $\theta=1.05^{\circ}$) for demonstration, and the coupling constant $g_0 D$ can then be greater than 1. The Coulomb repulsion can reduce $g_0$ to the weak coupling regime, where mean-field theory is applicable. In addition, because the band structure can be tuned by the twist angle and by hydrostatic pressure,  $g_0 D$ is less than one in a large parameter space defined by twist angle, chemical potential and pressure. 

If the dimensionless constant $g_0 D$ is indeed in the strong coupling regime, our weak coupling mean-field theory could break down, and other types of phonon driven instability should also be considered. Twisted bilayer graphene could be an ideal system to study the evolution from weak coupling to strong coupling regimes, because the bandwidth can be continuously tuned by pressure as demonstrated by the recent experimental work \cite{Dean2018Tuning}. Our theory provides a starting point to address this physics.

\section{Coherence length and modulation in the pair amplitude}
There are two different motions of the two paired electrons in one Cooper pair: one is the relative motion and the other is the center-of-mass motion. These two motions have different characteristic lengths in the superconducting {\it ground} state. 

For relative motion, the spatial separation between the two electrons can be roughly estimated by the coherence length $\xi_0 \sim \hbar v_F^*/E_g \sim (E_b/E_g) a_M$, where $v_F^*$ is the renormalized velocity of the nearly flat bands, $E_g$ is the superconducting gap in the quasiparticle energy spectrum, $a_M $ is the moir\'e period, and $E_b$ is the bandwidth of the flat bands. For superconducting transition temperature of 1.7 K, $E_g$ is about 0.25 meV. On the other hand, $E_b$ near magic angle is about few meV. Therefore, the coherence length is about few times of $a_M$. In the experiment \cite{Cao2018Super}, the coherence length has been found to be about 52 nm, which is about four times of the moir\'e period (13 nm) and is consistent with the above estimation. This consistency in turn shows that the Cooper pairing indeed forms between electrons in the flat bands.

On the other hand, the pair amplitude $\Delta(\rr)$, defined as $\langle \psi_\downarrow(\rr) \psi_\uparrow(\rr)\rangle$, corresponds to the center-of-mass motion. $\Delta(\rr)$ varies spatially with the moir\'e period $a_M$, and its spatial modulation follows the electron density variation in the normal state of the flat bands.

The above discussion is not unique to twisted bilayer graphene, but actually applies to all crystalline superconductors. We take superconducting aluminum as an example, which has a long coherence length $\xi_0 \approx 1600 $ nm. On the other hand, the pair amplitude $\Delta(\rr)=\langle \psi_\downarrow(\rr) \psi_\uparrow(\rr)\rangle$ in aluminum  has the lattice periodicity and varies within one unit cell following the variation of the electron wave function. This variation of the pair amplitude in the superconducting  {\it ground} state is typically ignored, because it is a modulation over a very short distance given by the lattice constant (0.4 nm in the case of aluminum). This modulation becomes dramatic in the moir\'e pattern because of the large lattice period of order 10 nm.

\end{document}